%% file: ms_v8.2.tex
\documentclass[letter]{aa} 
\usepackage{graphicx}
\usepackage{epstopdf}
\usepackage{txfonts}
\usepackage{natbib}
\usepackage{lscape}
\usepackage{balance}
\usepackage{longtable}
\newcommand{\lsim}{\ \raise -2.truept\hbox{\rlap{\hbox{$\sim$}}\raise
5.truept\hbox{$<$}\ }}
\newcommand{\gsim}{\ \raise -2.truept\hbox{\rlap{\hbox{$\sim$}}\raise
5.truept\hbox{$>$}\ }}

\newcommand{\kurt}{\hbox{\textit{kurt}}}
\newcommand\mM{\ifmmode(m{-}M)\else$(m{-}M)$\fi}
\newcommand\fred{\ensuremath{f_{\rm red}}}
\newcommand\zacs{\ensuremath{z_{F850LP}}}

\newcommand\gacs{\ensuremath{g_{F475W}}}

\begin{document}
   \title{Globular clusters of NGC\,3115 in the near--infrared
     \thanks{Based on observations made with ESO Telescopes at the La
       Silla Paranal Observatory under programme ID 60.A-9284. Tables
       \ref{tab_jhk}-\ref{tab_gmm_3g} are only available in electronic
       form via http://www.edpsciences.org}}

   \subtitle{Demonstrating the correctness of two opposing scenarios}

   \author{Michele Cantiello\inst{1}
          \and
   John P. Blakeslee\inst{2}
          \and
   Gabriella Raimondo\inst{1}
          \and
   Ana L. Chies-Santos\inst{3}
          \and
   Zachary G. Jennings\inst{4} 
          \and 
   Mark A. Norris\inst{5}
          \and 
   Harald Kuntschner\inst{6} }
\institute{INAF Osservatorio Astr. di Teramo, via Maggini, I-64100, Teramo, Italy  \email{cantiello@oa-teramo.inaf.it}
         \and
Dominion Astrophysical Observatory, Herzberg Institute of Astrophysics,
          NRC of Canada, Victoria, Canada 
          \and
School of Physics and Astronomy, University of Nottingham, University Park, Nottingham NG7 2RD, UK 
          \and
University of California Observatories, Santa Cruz, CA 95064, USA 
          \and
Max Planck Institut fur Astronomie, Konigstuhl 17, D-69117,
Heidelberg, Germany 
          \and
European Southern Observatory, Karl-Schwarzschild-Str. 2, D-85748
Garching bei M\"unchen, Germany 
 }
   \date{Received ---; accepted ---}

\authorrunning{Cantiello et al.}
\titlerunning{Reconciling GCs bimodality}

\abstract{We combined new near-infrared VLT/HAWK-I data of the
  globular clusters (GCs) in the isolated edge-on S0 galaxy \object{NGC\,3115}
  with optical and spectroscopic ones taken from the literature, with
  the aim of analyzing the multiband GC color distributions.  A recent
  study from the SLUGGS survey has shown that the GCs in this galaxy
  follow a bimodal distribution of Ca II triplet indices. Thus,
  \object{NGC\,3115} presents a critical example of a GC system with multiple,
  distinct, metallicity subpopulations, and this may argue against the
  ``projection'' scenario, which posits that the ubiquitous color
  bimodality mainly results from nonlinearities in the
  color-metallicity relations.  Using optical, NIR, and spectroscopic
  data, we found strong and consistent evidence of index bimodality,
  which independently confirms the metallicity bimodality in \object{NGC\,3115}
  GCs.  At the same time, we also found evidence for some color--color
  nonlinearity.  Taken in the broader context of previous studies, the
  multicolor consistency of the GC bimodality in \object{NGC\,3115} suggests
  that in cases where GC systems exhibit clear differences between
  their optical and optical--NIR color distributions (as in some giant
  ellipticals), the apparent inconsistencies most likely result from
  nonlinearities in the color--metallicity relations.}
   \keywords{galaxies: elliptical and lenticular, cD -- galaxies:
     individual (\object{NGC\,3115})-- galaxies: star clusters: general} \maketitle


\section{Introduction} 

The study of globular cluster (GC) systems in galaxies is one of the
keystones for understanding the processes at the base of the formation
and evolution of galaxies \citep{ashman92,brodie06}. Recently, the
interpretation of one of the most intriguing properties of GC systems
in early-type galaxies, the nearly universal presence of two distinct
peaks in the optical color distribution, has inspired a vigorous and
prolific debate
\citep[][]{yoon06,kundu07,yoon11b,chiessantos12,blake12a,usher12}.

The importance of GC bimodality was recognized before it was a
commonly observed property in early-type galaxies
\citep[ETGs,][]{schweizer87}. Historically, the bimodal GC color in
optical bands has been equated to metallicity ([Fe/H]) bimodality,
implying a fundamental constraint on GC and galaxy formation
scenarios. Metallicity bimodality requires two distinct epochs or
mechanisms of formation, or both, for the blue (metal-poor) and red
(metal-rich) GC subpopulations.

There are various proposed explanations for the GC color bimodality in
ETGs: {\it dissipational} merging of spirals, in which a merger-formed
population of red, metal-rich GCs is assumed to appear distinct from
the blue, metal-poor GCs of the progenitor spirals
\citep[][]{ashman92}; the {\it dry} hierarchical assembly, which
begins with a massive ``seed'' ETG that has a unimodal metal-rich GC
distribution, and in which it is possible to produce a bimodal
metallicity distribution through dissipationless accretion of many
early-type dwarfs \citep[][]{cote98}; and the {\it insitu} formation
scenario \citep{forbes97}. Most of these proposed mechanisms, though,
have assumed a simple linear conversion between [Fe/H] and color,
which seemed justified from the small fractional age variations among
the GCs \citep[][]{cohen98,kuntschner02,puzia05}.  However, this
assumption became the subject of debate when three independent works,
using observations and stellar population models, pointed out
non-negligible nonlinearities in the color-metallicity relations of
GCs \citep{peng06,richtler06,yoon06}. In particular, Yoon and
colleagues and \citeauthor{richtler06} demonstrated that these
nonlinearities naturally produce bimodal color histograms from
nonbimodal [Fe/H] distributions. This interpretation, dubbed the
projection effect, provided an alternative explanation based on
stellar evolution for the ubiquity of bimodal GC color distributions.
   
In this regard, \citet{cantiello07d} suggested the use of multicolor
GC histograms to verify the consistency of [Fe/H] distributions
derived from different colors. These authors highlighted the role of
optical to near-infrared (NIR) colors to distinguish between genuine
bimodality in [Fe/H] and projected bimodality in color.  If the
nonlinear projection is at work, then the [Fe/H] distributions
inferred from linear inversion of different color indices for the same
GC sample will show some degree of inconsistency (discordant [Fe/H]
peaks and/or fractions of GCs in each [Fe/H] component). The analysis
of optical-to-NIR GC colors in various ETGs, as well as $u$ to $z$
photometry in some Virgo cluster members, indicates that the nonlinear
projection effect is present at some level in these galaxies
\citep{blake12a,chiessantos12,yoon11b,yoon13}.

At the same time, the results of the SLUGGS
survey\footnote{\url{http://sluggs.ucolick.org.}}, which collected
spectra of $\sim1000$ GCs in 11 galaxies and derived [Fe/H] from the
calcium II triplet index, CaT, support true [Fe/H] bimodality in at
least some galaxies in addition to the \object{Milky Way}.  \citet{usher12}
found evidence for bimodal CaT distributions in six of eight galaxies
with sufficient numbers of GC spectra.  Nevertheless, the
spectroscopically and photometrically derived [Fe/H] distributions
show non-negligible differences in several galaxies of the SLUGGS
sample, thus lacking the aforementioned multi-index coherence.
However, the case of \object{\object{NGC\,3115}}, an isolated lenticular galaxy
at a distance of $\sim10$~Mpc \citep[][]{tonry01}, revealed highly
consistent [Fe/H] and color distributions, leading the authors to
present this galaxy as a critical test of [Fe/H] bimodality
\citep{brodie12}.  Previous optical VLT FORS2 spectroscopy for 17 GCs
in \object{NGC\,3115} showed hints of both a bimodal metallicity distribution
\textit{and} color or spectral index nonlinearity
\citep[][]{kuntschner02}, but it was limited by sample size.

In this Letter, we combine new NIR photometry with literature data to
investigate the consistency of the bimodality in the optical, NIR,
optical-NIR colors, and CaT, of the GCs in \object{NGC\,3115}.


\section{Observations}
\label{sec_obs}

To analyze the properties of the GC system in \object{NGC\,3115}, we took
advantage of NIR observations with VLT/HAWK-I, HST/ACS optical
photometry from \citet{jennings13} and spectroscopy from the SLUGGS
survey \citep{usher12}.

All photometry is corrected for Galactic foreground-reddening using
the \citet{sfd98} maps, with recalibration from \citet{sf11}. To
compare all data in the same photometric system, magnitudes were
converted to AB mag using the zero points derived using the SYNPHOT
task of IRAF/STSDAS and the spectrum of Vega from the
\citet{kurucz93b} atlas of stellar atmosphere models.

\subsection{VLT/HAWK-I near-infrared data}
We used HAWK-I science verification data. The observations were
carried out in very good seeing conditions, with
$\hbox{FWHM}\lsim0\arcsec.5$ in all three bands.  The science
observations centered on \object{NGC\,3115} were interleaved with sky exposures
of equal exposure time. The total on-source exposure times were 480s
in $J$, 960s in $H$, and 1620s in $Ks$.

A series of custom IDL scripts was used to carry out the data
reduction. Briefly, this consisted of subtracting dark exposures,
producing super-sky flats from the stacked, unregistered sky exposures,
and applying the derived flats to the individual science and sky exposures.
Finally, individual exposures were registered and coadded.

Because it is difficult to model the disk component of the galaxy, we
adopted the approach described in \citet{jordan04}. After
 taking the logarithm of the galaxy image, we fitted two-dimensional
 bicubic splines to the brightness distribution using SExtractor
 \citep{bertin96}. We then took the inverse logarithm of this model and subtracted it from the original image.

The photometry was then obtained from a second run of SExtractor on
the residual image. Because of the very strong SBF signal
\citep{tal90}, we added the galaxy model times a filter-dependent
constant to the SExtractor input $rms$ map to avoid detecting surface
brightness fluctuations as sources
\citep[see][]{jordan04}\footnote{The constants were chosen to minimize
  the number of false detections. However, spurious objects are also
  removed when matching the NIR with optical photometric and
  spectroscopic catalogs (Section \S\ref{sect_lit}).}.  To derive the
absolute calibration, we compared our photometry catalogs with the
2MASS point source catalog.  A total of 15 point sources in the $JH$
frames, and 11 in $Ks$ were used to derive the zero points, providing
$zp_{J}=25.75\pm0.06$ mag, $zp_{H}=26.03\pm0.06$ mag, and
$zp_{Ks}=25.03\pm0.09$ mag.  The list of matched sources is reported
in Table \ref{tab_jhk}.

\subsection{Optical photometry and spectroscopy}
\label{sect_lit}

We compared the NIR data with ACS optical photometry and with the
SLUGGS data.  The literature catalogs were selected on the basis of
either the radial velocity or the spatial extent of the source or for
both quantities. This allowed us to obtain GC catalogs with negligible
contamination from fore- or background sources.

The SLUGGS survey is based on Subaru Suprime-Cam $gi$ photometry and
Keck/Deimos Ca II triplet measurements. The final sample of matched
sources includes $N_{GC}=88$ GCs ({\it HS} sample hereafter).

The ACS catalog consists of deep $g_{F475W}$ and $z_{F850LP}$
photometry. We paired our NIR catalog with the ACS one, which produced
a sample of $N_{GC}=264$ GC candidates ({\it HA} sample).

Finally, we also considered the sample of objects obtained by
matching all three of the catalogs, which includes 74 GCs
({\it HSA} sample).

We analyzed the color and CaT distributions for the above three
samples using the Gaussian mixture modeling code
\citep[GMM,][]{muratov10}. This uses the likelihood-ratio test to
compare the goodness of fit for double Gaussians versus a single
Gaussian.  For the best-fit double model, it estimates the means and
widths of the two components, their separation DD in terms of combined
widths, and the kurtosis \kurt\ of the overall distribution. It also
provides uncertainties based on bootstrap resampling.

   \begin{figure*}
   \centering
   \includegraphics[width=0.35\textwidth]{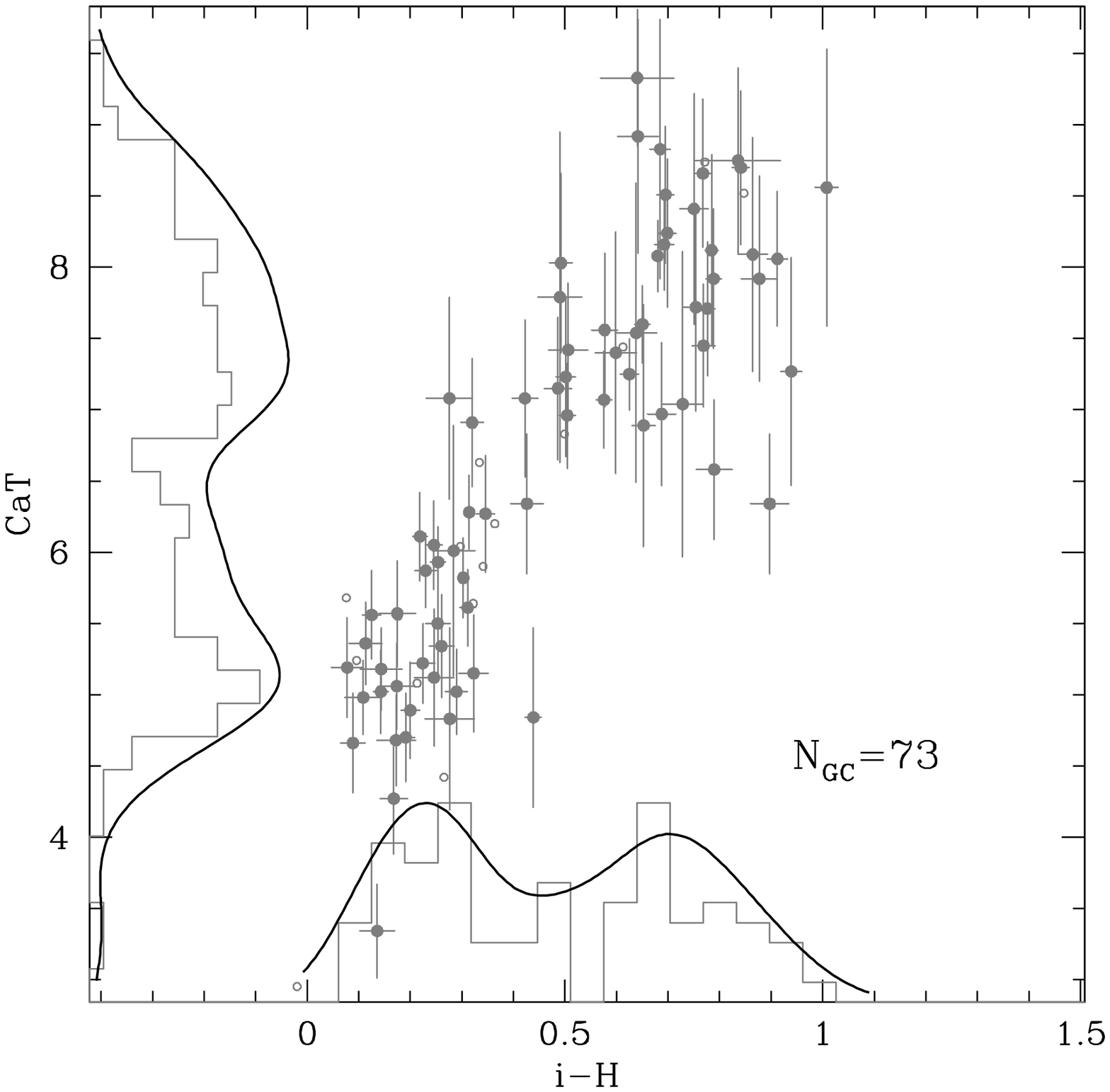}  \hskip-0.75 cm
   \includegraphics[width=0.35\textwidth]{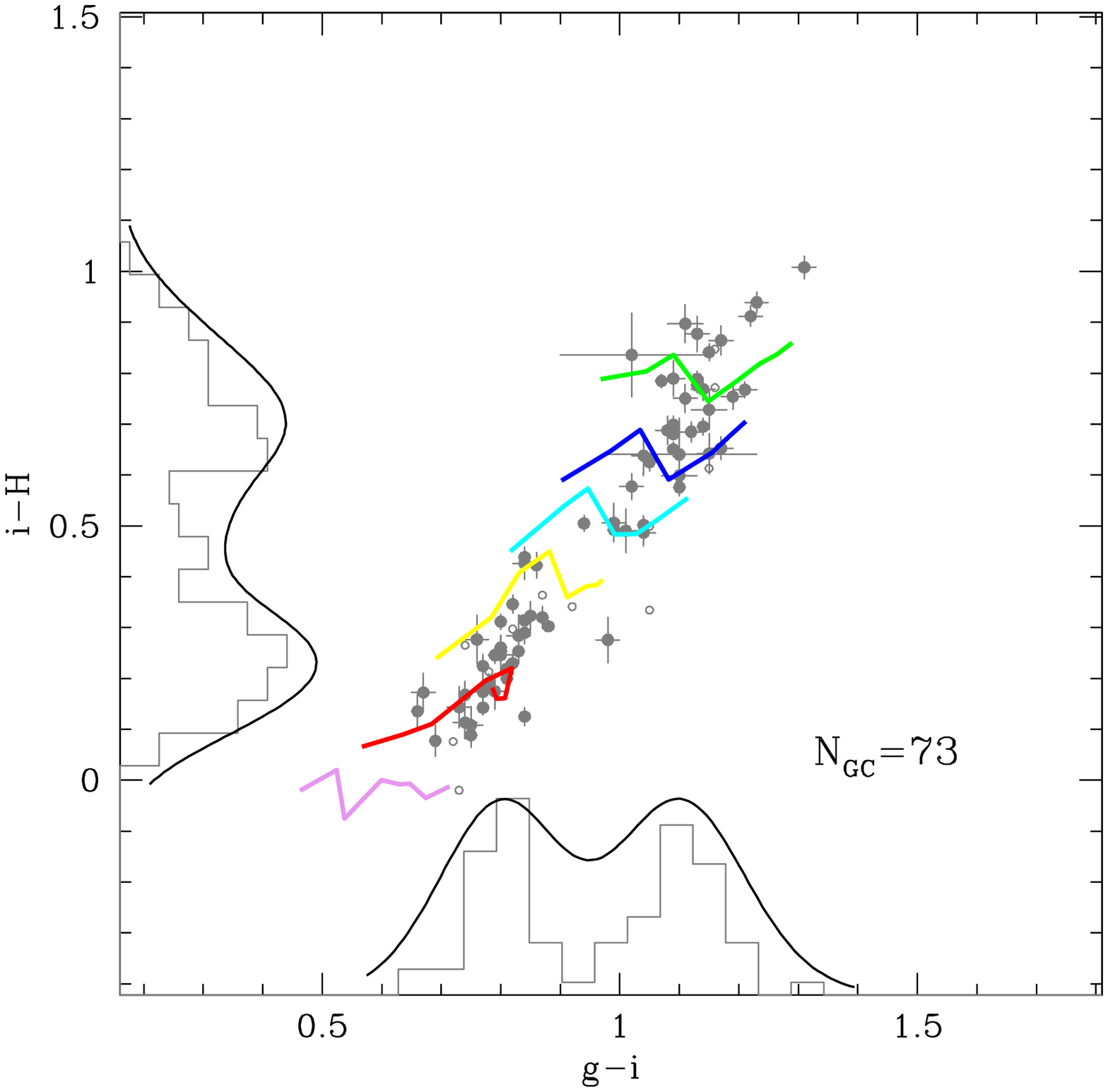}   \hskip-0.75 cm
   \includegraphics[width=0.35\textwidth]{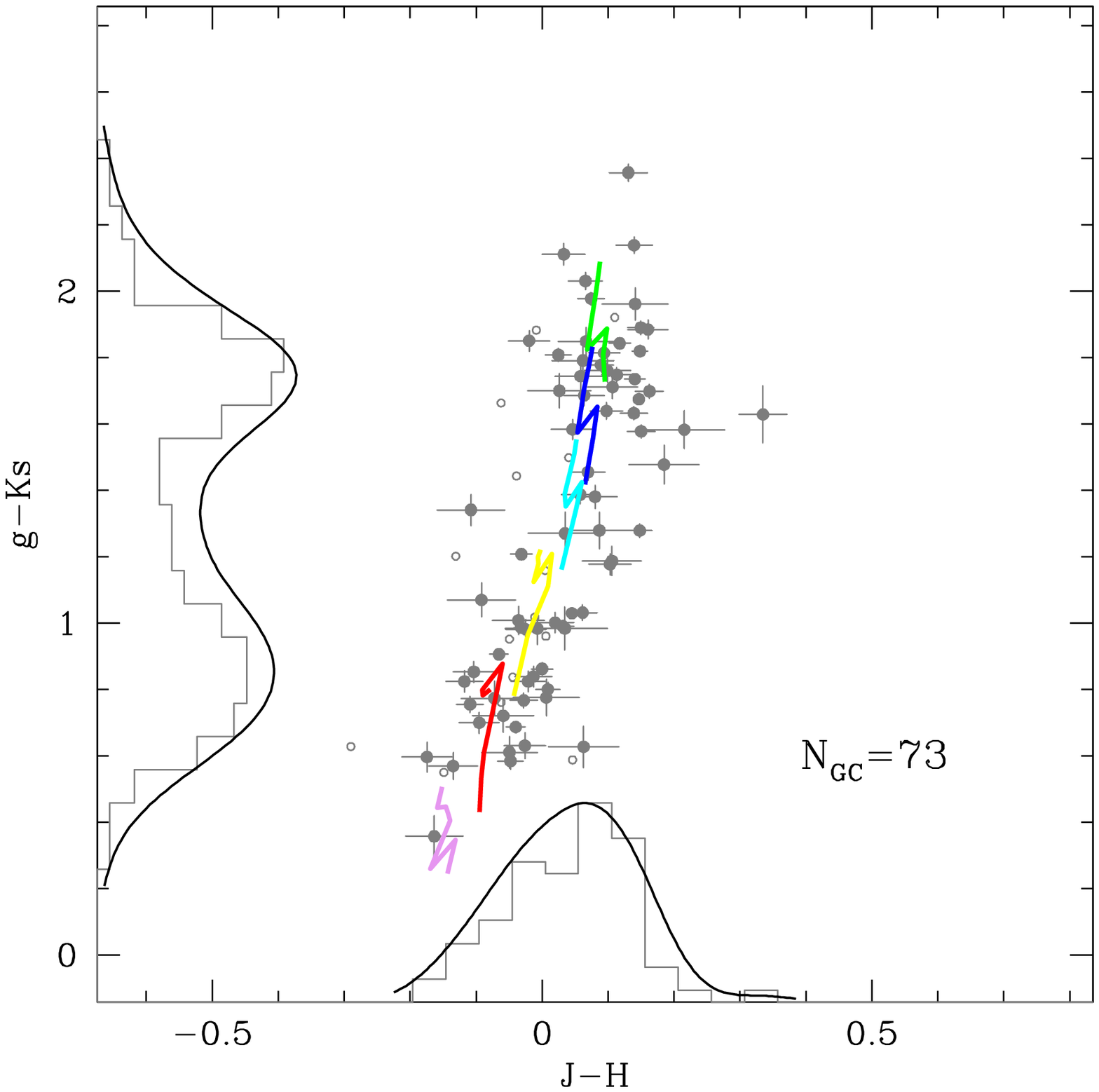}  \vskip-0.8 cm
    \caption{Color-CaT and a selection of color-color diagrams for the
      HS sample (sel\#1, full gray circles). Open circles mark GC
      candidates rejected by the adopted selection criteria. SPoT SSP
      models for [Fe/H]=$-2.3$ (pink), $-1.3$ (red), $-0.7$ (yellow),
      $-0.3$ (cyan), $0.0$ (blue) and $+0.3$ (green) dex are also
      shown. The age range is $2\leq t(Gyr)\leq14$, at fixed
      metallicity younger ages have bluer color. Histograms, and
      nonparametric density estimates for histograms are also shown
      with arbitrary scale. (See electronic version of the Journal for
      a colour version of the figure.)}
   \label{vltsluggs_sel1}
   \end{figure*}

\section{Analysis of the color distributions}
\label{sect_analysis}
 We analyzed the three samples mentioned in Section \S\ref{sect_lit}
 by adopting the two different selection criteria described below.
 First, to coherently compare our results with the recent analysis by
 \citet{brodie12}, we selected all GC candidates with CaT measurements
 with a signal-to-noise ratio $\,>12$ \AA$^{-1}$ (sel\#1
 hereafter\footnote{Since the HA sample lacks the CaT measurements
   used to define the sample sel\#1, in this case we chose to include
   only matching sources with $\Delta~mag\leq0.05$ in \textit{all}
   HAWK-I and ACS passbands.}). Second, for a consistent comparison of
 our analysis with a similar study of \object{NGC\,1399} GCs
 \citep{blake12a}, we selected only GCs within the magnitude range
 $20.5\leq m_i~(mag) \leq22.5$, and photometric uncertainty on color
 $\Delta~color\leq0.07$ mag (sel\#2).

The results of the GMM tests on the three GC samples with the two
different selection criteria are reported in Tables
\ref{tab_gmm}-\ref{tab_gmm2} and can be summarized as follows:

$i)$ HS sample (sel\#1) -- The color-color and color-CaT diagrams for
this sample and selection are shown in Figure \ref{vltsluggs_sel1}. In
general, the GMM results agree well for the various indices.  However,
for the purely NIR colors, the likelihood ratio test, $p(\chi^2)$, is
consistent with a bimodal distribution in $J{-}Ks$ and $H{-}Ks$, while
the $p$-values for the peak separation and kurtosis do not
significantly favor bimodality.

$ii)$ HA sample (sel\#1) -- Although the sample is nearly twice as
large as the HS sample, the relevance of color bimodality in optical
and optical-to-NIR data, and the lack of a strong evidence of
bimodality in NIR colors is nearly identical to the previous case.

$iii)$ HSA sample -- The sample, contains 58 GC candidates in the case
of sel\#1 and returns GMM values consistent with the two previous
samples (Figure \ref{alldata_sel1}).

$iv)$ sel\#1 versus sel\#2-- As expected, the results of GMM obtained
using sel\#1 or sel\#2 are very similar for the HS and HSA
samples. This is mostly a consequence of the small difference between
the catalogs obtained when adopting one selection or the
other. However, the similarity between selections \#1 and \#2 is also
true for the HA sample, which is twice as large as sel\#2, meaning
that it reaches $N_{GC}$ up to four times higher than some other
samples or selections.

$v)$ Although the values for the fraction of red GCs \fred\ derived
from different samples, or selection, or colors are largely
consistent, closer inspection shows the systematic nature of
\fred\ with respect to galactocentric distance $R_{GC}$, that is, the
radial gradient in mean GC color.  In particular, for sel\#1 we find
$\langle f_{\rm red}^{HA}\rangle=0.67\pm0.04$, while $\langle f_{\rm
  red}^{HS}\rangle=0.55\pm0.04$\footnote{For the selection \#2 we find
  $\langle f_{\rm red}^{HA}\rangle=0.56\pm0.09$, and $\langle f_{\rm
    red}^{HS}\rangle=0.52\pm0.06$.}.  Because the mean $R_{GC}$ of the
HA sample is smaller than that of the HS ($R_{GC}^{HA}\sim16\%$
smaller), the difference in \fred\ for the two samples reflects the
known tendency of red GCs to be more centrally concentrated than blue
ones \citep{kp97,larsen03}.

A summary of the Tables \ref{tab_gmm}-\ref{tab_gmm2} is reported in
Table \ref{tab_results}. The table gives the averaged $p$-values from
the different samples, obtained from coupling all the optical colors
(e.g. $g{-}i$; the column $N_{col}$ gives the number of colors used
for the average), all the optical-to-NIR (e.g. $g{-}H$), and all the
NIR colors (e.g. $J{-}Ks$). The most profound result here, again, is
the coherence between optical and optical-to-NIR colors, and the
already mentioned lack of bimodality in NIR colors based on peak
separation and kurtosis statistics.

Before moving on, we note that in some cases the GMM analysis
indicates that three Gaussians are preferred to (or as good as) the
double Gaussians model. The results of GMM fits with three components
are reported in Table \ref{tab_gmm_3g}.  Only cases where the least
populated peak contains at least $\sim10\%$ of the total population
are considered.  Although some colors for some samples are well fit
with three Gaussians, we stress the lack of coherence in these
tri-modal models for different colors.

\section{Comparison with population models}
\begin{figure}
   \centering
   \includegraphics[width=0.4\textwidth]{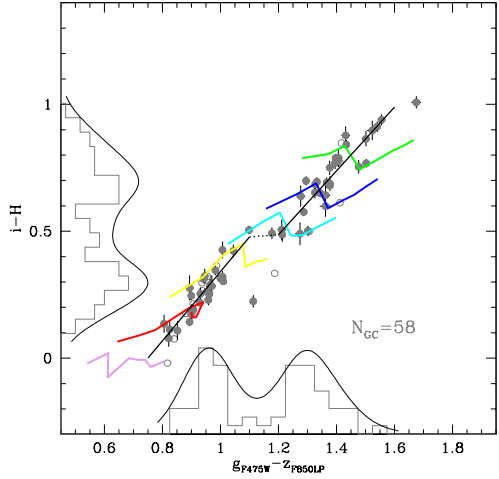} \vskip-0.5cm
   \caption{Similar to Figure \ref{vltsluggs_sel1}, but for the HSA
     sample.  The solid black lines show linear least-squares fits to
     the GCs with $\gacs{-}\zacs < 1.1$ and $1.2<\gacs{-}\zacs < 1.6$,
     respectively; the dotted nearly horizontal black line connects
     the endpoints of these fits. (See electronic version of the
     Journal for a colour version of the figure.)}
   \label{alldata_sel1}
   \end{figure}

Figures \ref{vltsluggs_sel1}-\ref{alldata_sel1} show the integrated
colors for simple stellar population (SSP) models from the SPoT group
\citep[][]{raimondo09}\footnote{\url{http://www.oa-teramo.inaf.it/spot}}.
The model grid ranges from 2 to 14 Gyr in SSP age, and from $-2.3$ to
$+0.3$ dex in [Fe/H].
Interpretation of integrated colors is hindered by the age-metallicity
degeneracy \citep{worthey94b}. In this case, though, the broad
wavelength coverage of the data and the higher sensitivity to [Fe/H]
of optical-to-NIR colors compared with purely optical colors allows
one to set more robust constraints on the stellar populations from the
full set of color-color diagrams. Overall, the SSP models and the data
agree well in general, and it is evident that the GC system covers
almost the full metallicity range of the models.

In Figure \ref{alldata_sel1} we show a color-color diagram for the HSA
sample (sel\#1). The sample from matching the three catalogs is
smaller than the others considered above. However, the features
observed in other samples are preserved, with lower scatter. Among
these we highlight first, the absence of a GC component with $t\leq8$
Gyr and [Fe/H]$\,\leq-2.3$ dex, or even the complete lack of clusters
with this or a lower metallicity. Second, on the high-metallicity
regime, some GC candidates ($\gacs{-}\zacs\geq1.5$ and $i{-}H\geq 0.9$
mag) are consistent with [Fe/H]$\geq+0.3$ dex.  The fraction of
high-metallicity GC candidates is $\sim5-10$\%.  Other color-color
diagrams also show a similar fraction of GCs above the [Fe/H]=0.3 dex
limit\footnote{Larger fractions, up to $\sim15\%$, are implied by the
  $z{-}J$ or $z{-}H$ colors. The data-to-model mismatch might in part
  result from the uncertainty in high-[Fe/H] SSP models for these
  colors that is caused by the partial overlap or proximity of the
  passbands and uncertainties in stellar atmosphere models in this
  wavelength regime.  However, a similar fraction of GCs with
  [Fe/H]$\gsim0.3$ dex was also found by \citet[][]{brodie12} -- we
  transformed their [Z/H] in [Fe/H] assuming the results of
  \citet{tantalo98}, with $[\alpha/Fe]=0.35$ \citep{pietrinferni06}.
  Moreover, these higher metal-rich fractions are for the HA sample,
  which is the most centrally concentrated of the considered samples
  (item $(iv)$, section \ref{sect_analysis}), and would therefore have
  the highest proportion of metal-rich GCs.}.

Furthermore, while the bluer GCs appear to match the locus of SSP
models for ages $t>10$ Gyr, the red clusters are more consistent with
$t<10$ Gyr. However, the age sensitivity of the models is weak, and
uncertainties in evolutionary {\it and} atmosphere models of AGB stars
-- such as low temperature opacity, the prescriptions for mass loss
and dust -- as well as the presence of extreme horizontal branch
stars, might lead to artificial age differences.  Nonetheless, the
suggested trend agrees with other studies that found the metal-poor
GCs to be 1–2 Gyr older than metal-rich ones
\citep[][]{puzia05,woodley10}.

An additional interesting feature of Figure \ref{alldata_sel1} is the
nonlinear shape of the color-color relation: at blue colors, $i{-}H$
increases steeply with $\gacs{-}\zacs$; the dependence flattens at
intermediate colors before becoming steep again.  This is similar to
the behavior found by \citet{blake12a} and resembles the
quasi-inflection point present in color-metallicity relations,
possibly indicating the shift in HB morphology around $i{-}H\sim$0.5
mag \citep{yoon06,cantiello07d}.

Finally, with regard to the dip in the color distributions, there is
general agreement among the various color indices for a deficit of GCs
in the $-1.2\lesssim[Fe/H]\lesssim -0.4$~dex range.  This confirms the
results from the CaT index \citep{brodie12} and indicates that CaT
provides metallicity estimates consistent with integrated colors (when
interpreted with the above models), at least at this intermediate
[Fe/H] regime.

\begin{table}
\caption{Average results of GMM tests}
\label{tab_results}
\centering
\scriptsize
\begin{tabular}{l c c c c c}
\hline\hline 
Col. range     &$ p(\chi^2)(\pm)$    &$ p(DD)(\pm)$        &$ p(kurt)(\pm)$        &\fred\ ($\pm$) &  $N_{cols}$   \\
                 &                              &                             &                                 &                       &       \\
\hline
\multicolumn{6}{c}{GC selection \#1} \\
Optical          &      0.00(0.02)    &     0.03(0.04)    &      0.00($<$0.01)  &     0.55(0.02)    &        5     \\       
Opt.+NIR         &      0.00(0.04)    &     0.15(0.04)    &      0.00(0.01)     &     0.56(0.02)    &       24     \\       
NIR              &      0.20(0.09)    &     0.67(0.11)    &      0.92(0.06)     &     0.44(0.12)    &        9     \\       
\hline
\multicolumn{6}{c}{GC selection \#2} \\
Optical          &      0.00(0.01)    &     0.02(0.05)    &      0.00($<$0.01)    &    0.56(0.02)    &        5     \\       
Opt.+NIR         &      0.00(0.03)    &     0.17(0.03)    &      0.00(0.01)     &    0.52(0.02)    &       24     \\       
NIR              &      0.07(0.11)    &     0.82(0.04)    &      0.94(0.07)     &    0.36(0.08)    &        9     \\       
\end{tabular} 
\end{table}

\section{Discussion and conclusions}

Our analysis of the GC system in \object{NGC\,3115} provides
additional definitive proof of GC [Fe/H] bimodality in this S0 galaxy,
the first to be firmly established beyond the \object{Milky Way}.
Various studies have noted that to rule out the possibility that
nonlinearities project a nonbimodal [Fe/H] distribution into a bimodal
color distribution, one must recover consistent [Fe/H] distributions
from multiple different photometric indices.  In particular,
\citet{cantiello07d} showed that optical-to-NIR colors are the most
useful in constraining the underlying metallicities.  This is in part
because the broad color baselines imply a lower sensitivity to the
detailed shape of the index-metallicity relation; for the converse
reason, bimodal [Fe/H] distributions may not be evident in the purely
NIR colors.  The present study confirms these expectations: $i)$ the
optical and optical-to-NIR colors are clearly bimodal and consistent
with each other in terms of proportions of red and blue GCs, and $ii)$
the pure NIR distributions give ambiguous results, in the sense that
different colors and/or statistical indicators indicate the presence
or lack of bimodality.

The comparison of GC colors with SSP models confirms earlier results
\citep{brodie12}, using CaT as a proxy for metallicity, which derived
a bimodal [Fe/H] distribution with a dip at $-1.3\lsim [Fe/H] \lsim
-0.3$ dex.  The model comparison also suggests that the metal-rich GCs
have slightly younger ages than the blue/metal-poor ones. Although
this conclusion is subject to model uncertainties, the size of the age
difference is consistent with that found between the metal-rich and
metal-poor GCs in the \object{Milky Way} \citep{vandenberg13}.

Some previous studies of optical-to-NIR GC colors in giant ellipticals
have found significantly different optical versus optical-to-NIR color
distributions.  For instance, using HST ACS and WFC3/IR data,
\citet{blake12a} showed that while the optical color distributions of
the Fornax giant elliptical \object{NGC\,1399} are clearly bimodal
\citep[supporting earlier results from][]{forte05}, the $V{-}H$ and
$I{-}H$ distributions are not, or they imply significantly
\textit{different} bimodal breakdowns than found for the optical
alone.  Similarly, \citet{chiessantos12} studied the $gzK$ color
distributions of 14 elliptical galaxies, mainly in the Virgo cluster,
and reported that double-peaked color distributions are more common in
$g{-}z$ than in the optical-NIR colors.  Both studies found
significant nonlinearity between the purely optical and optical-to-NIR
colors, and both concluded that bimodal optical color distributions
are not necessarily indicative of underlying bimodality in
metallicity.

The lack of consistency between the purely optical and optical--NIR
colors for \object{NGC\,1399} and some Virgo ellipticals contrasts
strongly with the coherent color bimodality observed in
\object{NGC\,3115}, for which the metallicities are also clearly
bimodal.  In light of this contrast, and with the evidence for
nonlinear color--color relations, the inconsistent optical and
optical-NIR color bimodalities in some galaxies imply that
nonlinearities do indeed play an important role in shaping the GC
color distributions in those galaxies.  Thus, while our multicolor
photometric analysis confirms the [Fe/H] bimodality of
\object{NGC\,3115} GCs, we conclude that the nonlinear projection
effect remains a viable explanation for the \textit{ubiquity} of
optical color bimodality and is the most likely cause in cases where
the optical-NIR colors lack obvious bimodality.

In summary: $i)$ optical and optical-to-NIR colors and CaT indices of
GCs in \object{NGC\,3115} are bimodal; $ii)$ the bimodal distributions derived
for different photometric and spectroscopic indices show good
consistency; $iii)$ evidence for bimodality is weak or absent for
purely NIR colors; $iv)$ our results agree with model predictions for
GC systems with truly bimodal [Fe/H] distributions, which provides
definitive proof of [Fe/H] bimodality in \object{NGC\,3115}, perhaps the first
galaxy beyond the Local Group for which this is the case; $v)$
comparison with SSP models confirms earlier metallicity results based
on CaT indices; $vi)$ despite the consistency of the color
distributions, we also observe color-color nonlinearities, most
clearly in the case of $i{-}H$ versus $g{-}z$.

Thus, the metallicity distributions of extragalactic GC systems and,
more specifically, the existence or lack of a universal bimodality of
[Fe/H] in ETGs, remains a matter of debate.  Indirectly, when taken in
the broader context of previous work, our analysis indicates that
optical GC color bimodalities have different causes in different
galaxies, with nonlinear color-metallicity relations playing an
important role for some previously studied giant ellipticals.
Finally, our study shows the effectiveness of optical-to-NIR colors 
as an unambiguous test for underlying metallicity bimodality.

\begin{acknowledgements}

Part of this work was supported by the FIRB-MIUR 2008 (P.I.
G. Imbriani).

\end{acknowledgements}

\bibliographystyle{aa}
\bibliography{cantiello_nov13}


\onltab{\input{table_jhk}}
\onltab{\input{table_gmm_v1}}
\onltab{\input{table_gmm2_v1}}
\onltab{\input{tab_gmm_3g}}

\end{document}

%% file: table_jhk.tex
\begin{table*}
\scriptsize
\caption{Matched HAWK--I $JHKs$ photometry for all detected sources}
\label{tab_jhk}
\centering
\begin{tabular}{l c c c c c c}
\hline\hline 
   ID  &   R.A.     &   Dec.     &            $m_J$         &           $m_H$          &            $m_{Ks}$      &$\langle CS\rangle$ \\
       & (J2000)    & (J2000)    &            (mag)         &           (mag)          &           (mag)        &           \\
   (1) &   (2)      &   (3)      &            (4)           &           (5)            &           (6)          &     (7)    \\
\hline
     1 & 151.261846 &  -7.778326 &     14.532 $\pm$   0.001 &    14.560 $\pm$   0.001 &    14.832 $\pm$   0.001 &      0.997 \\ 
     2 & 151.339249 &  -7.779547 &     15.262 $\pm$   0.002 &    15.290 $\pm$   0.001 &    15.713 $\pm$   0.001 &      0.990 \\ 
     3 & 151.328988 &  -7.781540 &     13.407 $\pm$   0.001 &    13.335 $\pm$   0.001 &    13.949 $\pm$   0.001 &      1.000 \\ 
     4 & 151.246152 &  -7.782556 &     20.196 $\pm$   0.013 &    20.526 $\pm$   0.018 &    21.869 $\pm$   0.061 &      0.020 \\ 
     5 & 151.280919 &  -7.782237 &     19.050 $\pm$   0.007 &    19.092 $\pm$   0.006 &    19.317 $\pm$   0.008 &      0.983 \\ 
\dots  & \dots      &  \dots     &     \dots                &     \dots               &   \dots                 &      \dots \\ 
\end{tabular} \\
\tablefoot{Columns list: (1) ID number; (2) right ascension; (3) declination; (4) $J-$band magnitude; (5) $H-$band magnitude; (6) $K_s-$band magnitude; (7) SExtractor mean $JHKs$ CLASS\_STAR parameter.}
\end{table*}

%% file: table_gmm_v1.tex
\begin{landscape}
\begin{table}
\scriptsize
\caption{Results of GMM runs on GCs selected using criterion \#1 (based on $\Delta CaT$, see text).}
\label{tab_gmm}
\centering
\begin{tabular}{cccccccccccc}
\hline\hline 
color    &         $p_1$      &  $p_2$            &  $\sigma_1$       &  $\sigma_2$      &$N_{GC}$&      frac$_2$  &   $p(\chi^2)$& $p(DD)$&$p(kurt)$&  DD      &   kurtosis  \\
     (1) &   (2)              &   (3)            &            (4)    &           (5)    &    (6)&        (7)    &       (8)     & (9)   & (10)    & (11)     &   (12)   \\
\hline
\multicolumn{12}{c}{HS (HAWK-I+SLUGGS) sample} \\
$g{-}i$  & 0.794 $\pm$ 0.010 & 1.103 $\pm$ 0.013 & 0.055 $\pm$ 0.008 & 0.075 $\pm$ 0.010 & 73 & 0.531 $\pm$ 0.067 & 0.001 & 0.023 & 0.001 & 4.71 $\pm$ 0.48 & -1.403 \\
$g{-}J$  & 1.051 $\pm$ 0.038 & 1.672 $\pm$ 0.053 & 0.104 $\pm$ 0.020 & 0.231 $\pm$ 0.032 & 73 & 0.561 $\pm$ 0.076 & 0.001 & 0.118 & 0.001 & 3.47 $\pm$ 0.51 & -1.241 \\
$g{-}H$  & 1.034 $\pm$ 0.028 & 1.806 $\pm$ 0.040 & 0.143 $\pm$ 0.020 & 0.202 $\pm$ 0.031 & 73 & 0.524 $\pm$ 0.070 & 0.001 & 0.033 & 0.001 & 4.41 $\pm$ 0.52 & -1.428 \\
$g{-}Ks$ & 0.874 $\pm$ 0.064 & 1.736 $\pm$ 0.069 & 0.225 $\pm$ 0.051 & 0.232 $\pm$ 0.059 & 73 & 0.505 $\pm$ 0.095 & 0.001 & 0.078 & 0.001 & 3.77 $\pm$ 0.49 & -1.273 \\
$i{-}J$  & 0.263 $\pm$ 0.030 & 0.567 $\pm$ 0.057 & 0.065 $\pm$ 0.020 & 0.163 $\pm$ 0.027 & 73 & 0.587 $\pm$ 0.125 & 0.001 & 0.453 & 0.012 & 2.44 $\pm$ 0.58 & -0.944 \\
$i{-}H$  & 0.219 $\pm$ 0.023 & 0.679 $\pm$ 0.037 & 0.078 $\pm$ 0.016 & 0.156 $\pm$ 0.025 & 73 & 0.568 $\pm$ 0.080 & 0.001 & 0.080 & 0.001 & 3.74 $\pm$ 0.55 & -1.364 \\
$i{-}Ks$ & 0.096 $\pm$ 0.077 & 0.647 $\pm$ 0.068 & 0.195 $\pm$ 0.051 & 0.161 $\pm$ 0.051 & 73 & 0.465 $\pm$ 0.137 & 0.025 & 0.219 & 0.009 & 3.08 $\pm$ 0.49 & -1.021 \\
$J{-}Ks$ &-0.164 $\pm$ 0.122 & 0.041 $\pm$ 0.056 & 0.157 $\pm$ 0.039 & 0.075 $\pm$ 0.031 & 73 & 0.364 $\pm$ 0.239 & 0.107 & 0.720 & 0.918 & 1.67 $\pm$ 0.88 & 0.536 \\
$H{-}Ks$ &-0.318 $\pm$ 0.118 &-0.112 $\pm$ 0.032 & 0.114 $\pm$ 0.035 & 0.088 $\pm$ 0.022 & 73 & 0.919 $\pm$ 0.240 & 0.228 & 0.623 & 0.976 & 2.03 $\pm$ 1.44 & 1.018 \\
$J{-}H$  &-0.099 $\pm$ 0.054 & 0.057 $\pm$ 0.090 & 0.046 $\pm$ 0.023 & 0.086 $\pm$ 0.030 & 73 & 0.889 $\pm$ 0.303 & 0.862 & 0.535 & 0.708 & 2.26 $\pm$ 1.05 & 0.012 \\
$CaT$    & 5.323 $\pm$ 0.254 & 7.705 $\pm$ 0.242 & 0.681 $\pm$ 0.230 & 0.745 $\pm$ 0.145 & 73 & 0.567 $\pm$ 0.110 & 0.010 & 0.156 & 0.005 & 3.34 $\pm$ 0.59 & -1.062 \\
\hline
\multicolumn{12}{c}{HA (HAWK-I+ACS) sample} \\
$g_{_{F475W}}{-}z_{_{F850LP}}$ & 0.947 $\pm$ 0.014 & 1.361 $\pm$ 0.016 & 0.070 $\pm$ 0.011 & 0.126 $\pm$ 0.012 & 128 & 0.664 $\pm$ 0.048 & 0.001 & 0.065 & 0.001 & 4.06 $\pm$ 0.35 & -1.254 \\
$g_{_{F475W}}{-}J$ & 1.072 $\pm$ 0.025 & 1.676 $\pm$ 0.032 & 0.099 $\pm$ 0.016 & 0.229 $\pm$ 0.022 & 128 & 0.681 $\pm$ 0.052 & 0.001 & 0.124 & 0.001 & 3.42 $\pm$ 0.35 & -1.145 \\
$g_{_{F475W}}{-}H$ & 1.067 $\pm$ 0.035 & 1.779 $\pm$ 0.034 & 0.145 $\pm$ 0.027 & 0.224 $\pm$ 0.022 & 128 & 0.654 $\pm$ 0.051 & 0.001 & 0.086 & 0.001 & 3.78 $\pm$ 0.31 & -1.147 \\
$g_{_{F475W}}{-}Ks$ & 0.869 $\pm$ 0.032 & 1.681 $\pm$ 0.035 & 0.153 $\pm$ 0.021 & 0.261 $\pm$ 0.028 & 128 & 0.679 $\pm$ 0.046 & 0.001 & 0.081 & 0.001 & 3.79 $\pm$ 0.35 & -1.063 \\
$z_{_{F850LP}}{-}J$ & 0.127 $\pm$ 0.057 & 0.329 $\pm$ 0.085 & 0.070 $\pm$ 0.027 & 0.120 $\pm$ 0.037 & 128 & 0.666 $\pm$ 0.268 & 0.099 & 0.546 & 0.012 & 2.06 $\pm$ 0.50 & -0.805 \\
$z_{_{F850LP}}{-}H$ & 0.119 $\pm$ 0.059 & 0.419 $\pm$ 0.024 & 0.089 $\pm$ 0.036 & 0.110 $\pm$ 0.026 & 128 & 0.640 $\pm$ 0.160 & 0.003 & 0.231 & 0.003 & 3.01 $\pm$ 0.72 & -0.947 \\
$z_{_{F850LP}}{-}Ks$ & -0.110 $\pm$ 0.061 & 0.299 $\pm$ 0.032 & 0.103 $\pm$ 0.037 & 0.163 $\pm$ 0.026 & 128 & 0.755 $\pm$ 0.114 & 0.011 & 0.201 & 0.038 & 3.00 $\pm$ 0.51 & -0.655 \\
$J{-}Ks$ & -0.031 $\pm$ 0.094 & -0.070 $\pm$ 0.183 & 0.047 $\pm$ 0.046 & 0.155 $\pm$ 0.052 & 128 & 0.820 $\pm$ 0.334 & 0.420 & 0.898 & 0.920 & 0.33 $\pm$ 1.81 & 0.475 \\
$H{-}Ks$ & -0.119 $\pm$ 0.048 & -0.073 $\pm$ 0.039 & 0.098 $\pm$ 0.024 & 0.028 $\pm$ 0.024 & 128 & 0.151 $\pm$ 0.261 & 0.435 & 0.863 & 0.569 & 0.64 $\pm$ 0.94 & -0.097 \\
$J{-}H$ & 0.041 $\pm$ 0.048 & 0.324 $\pm$ 0.130 & 0.079 $\pm$ 0.024 & 0.034 $\pm$ 0.043 & 128 & 0.029 $\pm$ 0.377 & 0.013 & 0.029 & 0.987 & 4.65 $\pm$ 1.92 & 1.102 \\
\hline
\multicolumn{12}{c}{HSA (HAWK-I+SLUGGS+ACS) sample} \\
$g_{_{F475W}}{-}z_{_{F850LP}}$ & 0.938 $\pm$ 0.016 & 1.372 $\pm$ 0.020 & 0.068 $\pm$ 0.012 & 0.119 $\pm$ 0.022 & 58 & 0.555 $\pm$ 0.075 & 0.001 & 0.036 & 0.001 & 4.46 $\pm$ 0.63 & -1.418 \\
$i{-}z_{_{F850LP}}$ & 0.139 $\pm$ 0.013 & 0.276 $\pm$ 0.013 & 0.040 $\pm$ 0.008 & 0.049 $\pm$ 0.009 & 58 & 0.596 $\pm$ 0.099 & 0.118 & 0.241 & 0.020 & 3.09 $\pm$ 0.44 & -0.949 \\ 
$g{-}i$ & 0.801 $\pm$ 0.030 & 1.114 $\pm$ 0.012 & 0.054 $\pm$ 0.031 & 0.074 $\pm$ 0.016 & 58 & 0.545 $\pm$ 0.085 & 0.001 & 0.013 & 0.001 & 4.86 $\pm$ 1.07 & -1.446 \\
$g_{_{F475W}}{-}J$ & 1.065 $\pm$ 0.029 & 1.687 $\pm$ 0.060 & 0.112 $\pm$ 0.022 & 0.254 $\pm$ 0.040 & 58 & 0.558 $\pm$ 0.088 & 0.001 & 0.215 & 0.002 & 3.17 $\pm$ 0.60 & -1.139 \\
$g_{_{F475W}}{-}H$ & 1.045 $\pm$ 0.037 & 1.794 $\pm$ 0.052 & 0.145 $\pm$ 0.026 & 0.239 $\pm$ 0.048 & 58 & 0.552 $\pm$ 0.080 & 0.001 & 0.097 & 0.001 & 3.79 $\pm$ 0.62 & -1.266 \\
$g_{_{F475W}}{-}Ks$ & 0.892 $\pm$ 0.059 & 1.741 $\pm$ 0.074 & 0.236 $\pm$ 0.047 & 0.268 $\pm$ 0.064 & 58 & 0.506 $\pm$ 0.098 & 0.014 & 0.176 & 0.008 & 3.36 $\pm$ 0.54 & -1.086 \\
$z_{_{F850LP}}{-}J$ & 0.127 $\pm$ 0.035 & 0.290 $\pm$ 0.087 & 0.038 $\pm$ 0.027 & 0.154 $\pm$ 0.043 & 58 & 0.657 $\pm$ 0.197 & 0.007 & 0.799 & 0.173 & 1.45 $\pm$ 1.08 & -0.658 \\
$z_{_{F850LP}}{-}H$ & 0.133 $\pm$ 0.049 & 0.457 $\pm$ 0.056 & 0.103 $\pm$ 0.030 & 0.106 $\pm$ 0.030 & 58 & 0.454 $\pm$ 0.154 & 0.055 & 0.267 & 0.010 & 3.10 $\pm$ 0.66 & -1.048 \\
$z_{_{F850LP}}{-}Ks$ & 0.042 $\pm$ 0.143 & 0.431 $\pm$ 0.133 & 0.227 $\pm$ 0.068 & 0.150 $\pm$ 0.059 & 58 & 0.260 $\pm$ 0.280 & 0.860 & 0.669 & 0.252 & 2.03 $\pm$ 0.62 & -0.577 \\
$g{-}J$ & 1.069 $\pm$ 0.025 & 1.714 $\pm$ 0.044 & 0.107 $\pm$ 0.017 & 0.223 $\pm$ 0.035 & 58 & 0.554 $\pm$ 0.078 & 0.001 & 0.111 & 0.001 & 3.68 $\pm$ 0.61 & -1.315 \\
$g{-}H$ & 1.045 $\pm$ 0.029 & 1.814 $\pm$ 0.040 & 0.134 $\pm$ 0.020 & 0.214 $\pm$ 0.034 & 58 & 0.556 $\pm$ 0.072 & 0.001 & 0.049 & 0.001 & 4.30 $\pm$ 0.63 & -1.416 \\
$g{-}Ks$ & 0.903 $\pm$ 0.068 & 1.769 $\pm$ 0.067 & 0.243 $\pm$ 0.050 & 0.230 $\pm$ 0.063 & 58 & 0.498 $\pm$ 0.096 & 0.004 & 0.116 & 0.001 & 3.66 $\pm$ 0.56 & -1.230 \\
$i{-}J$ & 0.302 $\pm$ 0.027 & 0.661 $\pm$ 0.062 & 0.087 $\pm$ 0.023 & 0.124 $\pm$ 0.033 & 58 & 0.424 $\pm$ 0.142 & 0.001 & 0.168 & 0.009 & 3.35 $\pm$ 0.80 & -1.064 \\
$i{-}H$ & 0.235 $\pm$ 0.026 & 0.686 $\pm$ 0.040 & 0.081 $\pm$ 0.018 & 0.159 $\pm$ 0.027 & 58 & 0.589 $\pm$ 0.077 & 0.001 & 0.127 & 0.001 & 3.57 $\pm$ 0.62 & -1.335 \\
$i{-}Ks$ & 0.111 $\pm$ 0.084 & 0.659 $\pm$ 0.089 & 0.203 $\pm$ 0.060 & 0.170 $\pm$ 0.056 & 58 & 0.460 $\pm$ 0.159 & 0.122 & 0.295 & 0.017 & 2.94 $\pm$ 0.56 & -0.983 \\
$J{-}Ks$ & -0.169 $\pm$ 0.146 & 0.030 $\pm$ 0.052 & 0.159 $\pm$ 0.049 & 0.076 $\pm$ 0.030 & 58 & 0.394 $\pm$ 0.249 & 0.166 & 0.778 & 0.963 & 1.60 $\pm$ 1.23 & 0.874 \\
$H{-}Ks$ & -0.500 $\pm$ 0.118 & -0.131 $\pm$ 0.055 & 0.023 $\pm$ 0.033 & 0.090 $\pm$ 0.025 & 58 & 0.983 $\pm$ 0.321 & 0.088 & 0.004 & 0.989 & 5.62 $\pm$ 1.57 & 1.380 \\
$J{-}H$ & -0.002 $\pm$ 0.079 & 0.098 $\pm$ 0.036 & 0.083 $\pm$ 0.029 & 0.054 $\pm$ 0.020 & 58 & 0.480 $\pm$ 0.297 & 0.475 & 0.800 & 0.584 & 1.44 $\pm$ 1.15 & -0.194 \\
$CaT$ & 5.693 $\pm$ 0.333 & 7.833 $\pm$ 0.290 & 0.836 $\pm$ 0.279 & 0.618 $\pm$ 0.149 & 58 & 0.504 $\pm$ 0.159 & 0.044 & 0.303 & 0.044 & 2.91 $\pm$ 0.67 & -0.856 \\
\end{tabular} \\
\tablefoot{Columns list: (1) color; (2-3) mean and uncertainty of the
  first and second peaks in the double-Gaussian model; (4-5) width and
  uncertainty of the first and second peaks; (6) number of GC
  candidates selected; (7) fraction of GC candidates associated with
  the second, red, peak; (8-10) GMM $p$-values based on the likelihood-ratio test $p(\chi^2)$, peak separation $p(DD)$, and kurtosis
  $p(kurt)$, indicating the significance of the preference for a
  double-Gaussian over a single-Gaussian model (lower $p$-values are
  more significant); (11) separation of the peaks in units of the two
  Gaussian widths; (12) kurtosis of the distribution (DD$ \geq$2 and
  negative kurtosis are required for significative split between the
  two Gaussian distributions).}
\end{table}
\end{landscape}

%% file: table_gmm2_v1.tex
\begin{landscape}
\begin{table}
\scriptsize
\caption{Results of GMM runs on GCs selected using criterion \#2 (based on $\Delta~color$ and magnitude, see text).}
\label{tab_gmm2}
\centering
\begin{tabular}{cccccccccccc}
\hline\hline 
color    &         $p_1$      &  $p_2$            &  $\sigma_1$       &  $\sigma_2$      &$N_{GC}$&      frac$_2$  &   $p(\chi^2)$& $p(DD)$&$p(kurt)$&  DD      &   kurtosis  \\
     (1) &   (2)              &   (3)            &            (4)    &           (5)    &    (6)&        (7)    &       (8)     & (9)   & (10)    & (11)     &   (12)   \\
\hline
\multicolumn{12}{c}{HS (HAWK-I+SLUGGS) sample} \\
$g{-}i$ & 0.783 $\pm$ 0.011 & 1.106 $\pm$ 0.013 & 0.060 $\pm$ 0.007 & 0.076 $\pm$ 0.011 & 68 & 0.521 $\pm$ 0.064 & 0.001 & 0.023 & 0.001 & 4.74 $\pm$ 0.54 & -1.467 \\
$g{-}J$ & 1.032 $\pm$ 0.083 & 1.672 $\pm$ 0.110 & 0.112 $\pm$ 0.057 & 0.251 $\pm$ 0.066 & 68 & 0.566 $\pm$ 0.144 & 0.001 & 0.167 & 0.001 & 3.30 $\pm$ 0.59 & -1.292 \\
$g{-}H$ & 1.014 $\pm$ 0.050 & 1.804 $\pm$ 0.067 & 0.167 $\pm$ 0.033 & 0.221 $\pm$ 0.051 & 68 & 0.500 $\pm$ 0.085 & 0.001 & 0.062 & 0.001 & 4.04 $\pm$ 0.57 & -1.389 \\
$g{-}Ks$ & 0.867 $\pm$ 0.083 & 1.710 $\pm$ 0.084 & 0.221 $\pm$ 0.058 & 0.261 $\pm$ 0.068 & 65 & 0.536 $\pm$ 0.112 & 0.003 & 0.134 & 0.003 & 3.49 $\pm$ 0.54 & -1.210 \\
$i{-}J$ & 0.308 $\pm$ 0.040 & 0.671 $\pm$ 0.072 & 0.101 $\pm$ 0.025 & 0.120 $\pm$ 0.034 & 69 & 0.375 $\pm$ 0.146 & 0.002 & 0.190 & 0.008 & 3.28 $\pm$ 0.66 & -0.994 \\
$i{-}H$ & 0.219 $\pm$ 0.034 & 0.697 $\pm$ 0.047 & 0.103 $\pm$ 0.022 & 0.155 $\pm$ 0.029 & 69 & 0.510 $\pm$ 0.084 & 0.001 & 0.114 & 0.001 & 3.63 $\pm$ 0.49 & -1.309 \\
$i{-}Ks$ & 0.068 $\pm$ 0.088 & 0.608 $\pm$ 0.085 & 0.189 $\pm$ 0.051 & 0.191 $\pm$ 0.055 & 69 & 0.502 $\pm$ 0.155 & 0.114 & 0.322 & 0.008 & 2.84 $\pm$ 0.43 & -0.997 \\
$J{-}Ks$ & -0.234 $\pm$ 0.141 & -0.028 $\pm$ 0.051 & 0.173 $\pm$ 0.059 & 0.104 $\pm$ 0.029 & 60 & 0.686 $\pm$ 0.240 & 0.100 & 0.798 & 0.977 & 1.45 $\pm$ 1.26 & 1.063 \\
$H{-}Ks$ & -0.147 $\pm$ 0.107 & -0.074 $\pm$ 0.025 & 0.124 $\pm$ 0.046 & 0.031 $\pm$ 0.023 & 61 & 0.368 $\pm$ 0.210 & 0.013 & 0.845 & 0.988 & 0.82 $\pm$ 1.57 & 1.381 \\
$J{-}H$ & -0.128 $\pm$ 0.054 & 0.037 $\pm$ 0.095 & 0.027 $\pm$ 0.023 & 0.092 $\pm$ 0.028 & 67 & 0.919 $\pm$ 0.294 & 0.781 & 0.474 & 0.729 & 2.43 $\pm$ 1.01 & 0.019 \\
$CaT$ & 5.246 $\pm$ 0.256 & 7.707 $\pm$ 0.257 & 0.828 $\pm$ 0.243 & 0.714 $\pm$ 0.129 & 69 & 0.523 $\pm$ 0.106 & 0.022 & 0.217 & 0.015 & 3.18 $\pm$ 0.51 & -0.926 \\
\hline
\multicolumn{12}{c}{HA (HAWK-I+ACS) sample} \\
$g_{_{F475W}}{-}z_{_{F850LP}}$ & 0.931 $\pm$ 0.009 & 1.325 $\pm$ 0.018 & 0.060 $\pm$ 0.006 & 0.149 $\pm$ 0.010 & 252 & 0.596 $\pm$ 0.037 & 0.001 & 0.118 & 0.001 & 3.47 $\pm$ 0.29 & -1.307 \\
$g_{_{F475W}}{-}J$ & 1.089 $\pm$ 0.048 & 1.699 $\pm$ 0.061 & 0.173 $\pm$ 0.056 & 0.214 $\pm$ 0.036 & 207 & 0.500 $\pm$ 0.097 & 0.001 & 0.127 & 0.002 & 3.14 $\pm$ 0.38 & -0.711 \\
$g_{_{F475W}}{-}H$ & 1.034 $\pm$ 0.028 & 1.760 $\pm$ 0.032 & 0.170 $\pm$ 0.020 & 0.231 $\pm$ 0.021 & 215 & 0.563 $\pm$ 0.045 & 0.001 & 0.102 & 0.001 & 3.58 $\pm$ 0.24 & -1.188 \\
$g_{_{F475W}}{-}Ks$ & 0.837 $\pm$ 0.025 & 1.670 $\pm$ 0.039 & 0.168 $\pm$ 0.028 & 0.300 $\pm$ 0.045 & 165 & 0.674 $\pm$ 0.050 & 0.001 & 0.118 & 0.038 & 3.43 $\pm$ 0.43 & -0.573 \\
$z_{_{F850LP}}{-}J$ & 0.148 $\pm$ 0.033 & 0.364 $\pm$ 0.083 & 0.089 $\pm$ 0.028 & 0.122 $\pm$ 0.036 & 207 & 0.432 $\pm$ 0.238 & 0.003 & 0.424 & 0.183 & 2.03 $\pm$ 0.71 & -0.348 \\
$z_{_{F850LP}}{-}H$ & 0.121 $\pm$ 0.037 & 0.442 $\pm$ 0.042 & 0.133 $\pm$ 0.028 & 0.122 $\pm$ 0.029 & 215 & 0.465 $\pm$ 0.136 & 0.021 & 0.282 & 0.022 & 2.51 $\pm$ 0.39 & -0.586 \\
$z_{_{F850LP}}{-}Ks$ & -0.080 $\pm$ 0.059 & 0.330 $\pm$ 0.028 & 0.150 $\pm$ 0.035 & 0.155 $\pm$ 0.025 & 168 & 0.645 $\pm$ 0.115 & 0.011 & 0.254 & 0.044 & 2.69 $\pm$ 0.48 & -0.562 \\
$J{-}Ks$ & -0.086 $\pm$ 0.090 & -0.018 $\pm$ 0.179 & 0.160 $\pm$ 0.044 & 0.055 $\pm$ 0.053 & 142 & 0.220 $\pm$ 0.312 & 0.255 & 0.845 & 0.909 & 0.57 $\pm$ 1.78 & 0.408 \\
$H{-}Ks$ & -0.128 $\pm$ 0.030 & -0.072 $\pm$ 0.060 & 0.118 $\pm$ 0.029 & 0.031 $\pm$ 0.042 & 148 & 0.218 $\pm$ 0.203 & 0.046 & 0.845 & 0.983 & 0.65 $\pm$ 0.86 & 1.045 \\
$J{-}H$ & 0.055 $\pm$ 0.092 & 0.030 $\pm$ 0.088 & 0.061 $\pm$ 0.031 & 0.110 $\pm$ 0.032 & 169 & 0.641 $\pm$ 0.319 & 0.577 & 0.908 & 0.904 & 0.28 $\pm$ 1.58 & 0.412 \\
\hline
\multicolumn{12}{c}{HSA (HAWK-I+SLUGGS+ACS) sample} \\
$g_{_{F475W}}{-}z_{_{F850LP}}$ & 0.917 $\pm$ 0.016 & 1.367 $\pm$ 0.025 & 0.065 $\pm$ 0.011 & 0.120 $\pm$ 0.022 & 56 & 0.573 $\pm$ 0.073 & 0.001 & 0.021 & 0.001 & 4.65 $\pm$ 0.68 & -1.453 \\
$i{-}z_{_{F850LP}}$ & 0.140 $\pm$ 0.017 & 0.274 $\pm$ 0.018 & 0.030 $\pm$ 0.010 & 0.054 $\pm$ 0.011 & 56 & 0.630 $\pm$ 0.127 & 0.032 & 0.277 & 0.005 & 3.06 $\pm$ 0.59 & -1.129 \\ 
$g_{_{F475W}}{-}i$ & 0.789 $\pm$ 0.027 & 1.116 $\pm$ 0.014 & 0.057 $\pm$ 0.025 & 0.071 $\pm$ 0.014 & 56 & 0.554 $\pm$ 0.082 & 0.001 & 0.008 & 0.001 & 5.07 $\pm$ 0.91 & -1.477 \\
$g_{_{F475W}}{-}J$ & 1.052 $\pm$ 0.037 & 1.705 $\pm$ 0.069 & 0.124 $\pm$ 0.023 & 0.248 $\pm$ 0.042 & 56 & 0.554 $\pm$ 0.093 & 0.001 & 0.202 & 0.002 & 3.32 $\pm$ 0.63 & -1.226 \\
$g_{_{F475W}}{-}H$ & 1.019 $\pm$ 0.044 & 1.807 $\pm$ 0.058 & 0.168 $\pm$ 0.028 & 0.229 $\pm$ 0.049 & 56 & 0.535 $\pm$ 0.080 & 0.001 & 0.074 & 0.001 & 3.92 $\pm$ 0.62 & -1.294 \\
$g_{_{F475W}}{-}Ks$ & 0.835 $\pm$ 0.099 & 1.708 $\pm$ 0.104 & 0.245 $\pm$ 0.059 & 0.285 $\pm$ 0.077 & 56 & 0.542 $\pm$ 0.124 & 0.027 & 0.208 & 0.005 & 3.28 $\pm$ 0.50 & -1.132 \\
$z_{_{F850LP}}{-}J$ & 0.182 $\pm$ 0.026 & 0.491 $\pm$ 0.061 & 0.105 $\pm$ 0.022 & 0.048 $\pm$ 0.034 & 56 & 0.185 $\pm$ 0.127 & 0.011 & 0.098 & 0.106 & 3.78 $\pm$ 0.84 & -0.756 \\
$z_{_{F850LP}}{-}H$ & 0.114 $\pm$ 0.053 & 0.461 $\pm$ 0.059 & 0.114 $\pm$ 0.031 & 0.104 $\pm$ 0.032 & 56 & 0.438 $\pm$ 0.160 & 0.049 & 0.244 & 0.008 & 3.19 $\pm$ 0.54 & -1.096 \\
$z_{_{F850LP}}{-}Ks$ & -0.241 $\pm$ 0.153 & 0.222 $\pm$ 0.134 & 0.135 $\pm$ 0.073 & 0.228 $\pm$ 0.066 & 56 & 0.810 $\pm$ 0.287 & 0.718 & 0.498 & 0.175 & 2.47 $\pm$ 0.62 & -0.675 \\
$g{-}J$ & 1.056 $\pm$ 0.045 & 1.717 $\pm$ 0.066 & 0.119 $\pm$ 0.025 & 0.233 $\pm$ 0.041 & 56 & 0.571 $\pm$ 0.090 & 0.001 & 0.135 & 0.001 & 3.57 $\pm$ 0.61 & -1.365 \\
$g{-}H$ & 1.028 $\pm$ 0.057 & 1.818 $\pm$ 0.062 & 0.170 $\pm$ 0.040 & 0.222 $\pm$ 0.051 & 56 & 0.548 $\pm$ 0.091 & 0.001 & 0.063 & 0.001 & 4.00 $\pm$ 0.61 & -1.393 \\
$g{-}Ks$ & 0.913 $\pm$ 0.109 & 1.759 $\pm$ 0.079 & 0.262 $\pm$ 0.066 & 0.243 $\pm$ 0.062 & 53 & 0.540 $\pm$ 0.112 & 0.019 & 0.187 & 0.006 & 3.35 $\pm$ 0.60 & -1.136 \\
$i{-}J$ & 0.314 $\pm$ 0.036 & 0.697 $\pm$ 0.069 & 0.098 $\pm$ 0.023 & 0.105 $\pm$ 0.039 & 56 & 0.392 $\pm$ 0.144 & 0.001 & 0.099 & 0.003 & 3.77 $\pm$ 0.84 & -1.152 \\
$i{-}H$ & 0.258 $\pm$ 0.053 & 0.728 $\pm$ 0.061 & 0.130 $\pm$ 0.035 & 0.142 $\pm$ 0.038 & 56 & 0.494 $\pm$ 0.116 & 0.009 & 0.168 & 0.001 & 3.45 $\pm$ 0.59 & -1.283 \\
$i{-}Ks$ & 0.186 $\pm$ 0.134 & 0.687 $\pm$ 0.108 & 0.248 $\pm$ 0.077 & 0.168 $\pm$ 0.067 & 53 & 0.401 $\pm$ 0.220 & 0.419 & 0.561 & 0.075 & 2.36 $\pm$ 0.56 & -0.836 \\
$J{-}Ks$ & -0.242 $\pm$ 0.113 & -0.013 $\pm$ 0.040 & 0.178 $\pm$ 0.055 & 0.094 $\pm$ 0.025 & 47 & 0.661 $\pm$ 0.212 & 0.049 & 0.783 & 0.981 & 1.61 $\pm$ 1.08 & 1.141 \\
$H{-}Ks$ & -0.155 $\pm$ 0.093 & -0.075 $\pm$ 0.022 & 0.124 $\pm$ 0.032 & 0.031 $\pm$ 0.020 & 49 & 0.346 $\pm$ 0.199 & 0.044 & 0.889 & 0.982 & 0.89 $\pm$ 1.23 & 1.256 \\
$J{-}H$ & -0.001 $\pm$ 0.069 & 0.104 $\pm$ 0.038 & 0.084 $\pm$ 0.028 & 0.051 $\pm$ 0.019 & 54 & 0.313 $\pm$ 0.301 & 0.770 & 0.787 & 0.379 & 1.51 $\pm$ 0.99 & -0.437 \\
$CaT$ & 5.661 $\pm$ 0.429 & 7.904 $\pm$ 0.325 & 1.091 $\pm$ 0.347 & 0.635 $\pm$ 0.168 & 56 & 0.447 $\pm$ 0.148 & 0.071 & 0.502 & 0.157 & 2.51 $\pm$ 0.70 & -0.692 \\
\end{tabular} \\
\tablefoot{Columns list: same as in Table \ref{tab_gmm}}
\end{table}
\end{landscape}

%% file: tab_gmm_3g.tex
\begin{landscape}
\begin{table}
\scriptsize
\caption{Results of GMM runs for the three Gaussian model with GCs selected using criterion \#1.}
\label{tab_gmm_3g}
\centering
\begin{tabular}{ccccccccc}
\hline\hline 
color     &$p_1$ ($N_{GC}$, $\sigma$)     &        $p_2$($N_{GC}$, $\sigma$)&        $p_3$($N_{GC}$, $\sigma$)&      $p(\chi^2)$   &       $p(DD)$   &      $p(kurt)$   &  DD          &      kurtosis     \\
     (1)  &   (2)                       &   (3)                         &            (4)               &           (5)     &    (6)           &        (7)       &       (8)        & (9)           \\
\hline
\multicolumn{9}{c}{HS (HAWK-I+SLUGGS) sample ($N_{GC}^{tot}=73$)} \\
$g{-}J$      &       1.063(34.5, 0.114)      &       1.472(6.5, 0.049)      &       1.746(32.0, 0.197)       &       0.001      &       0.033      &       0.001      &     4.67       &       -1.241    \\
$i{-}J$      &       0.251(25.9, 0.058)      &       0.428(25.5, 0.112)      &       0.685(21.5, 0.111)      &       0.003      &       0.718      &       0.012      &     1.99       &       -0.944    \\
$i{-}H$      &       0.221(32.4, 0.078)      &       0.475(7.1, 0.033)      &       0.732(33.5, 0.120)      &       0.001      &       0.098      &       0.001      &      4.23       &       -1.364    \\
$CaT$        &       5.064(13.1, 0.285)      &       6.162(29.8, 1.141)      &       7.881(30.1, 0.681)      &       0.019      &       0.89      &       0.005      &      1.32       &       -1.062    \\
\hline
\multicolumn{9}{c}{HA (HAWK-I+ACS) sample  ($N_{GC}^{tot}=128$)} \\
$g{-}z$      &       0.951(44.6, 0.073)      &       1.177(8.5, 0.027)      &       1.388(74.8, 0.107)      &       0.001      &       0.118      &       0.001      &         4.11   &    -1.254    \\
$g{-}J$      &       1.078(43.1, 0.102)      &       1.412(8.1, 0.060)      &       1.718(76.8, 0.206)      &       0.001      &       0.136      &       0.001      &         4.00   &    -1.145    \\
$g{-}Ks$     &       0.878(42.6, 0.159)      &       1.636(73.1, 0.215)      &       2.023(12.2, 0.185)      &       0.001      &       0.11      &       0.001      &         4.01   &    -1.063    \\
$z{-}J$      &       0.123(45.7, 0.068)      &       0.313(72.5, 0.094)      &       0.521(9.8, 0.024)      &       0.036      &       0.444      &       0.012      &         2.32   &    -0.805    \\
$z{-}H$      &       0.110(44.2, 0.081)      &       0.264(8.9, 0.025)      &       0.435(74.9, 0.099)      &       0.01      &       0.449      &       0.003      &          2.58   &    -0.947    \\
\hline
\multicolumn{9}{c}{HSA (HAWK-I+SLUGGS+ACS) sample  ($N_{GC}^{tot}=58$)} \\
$g_{_{F475W}}{-}J$      &       1.069(26.6, 0.114)      &       1.629(25.6, 0.191)      &       2.035(5.8, 0.112)      &       0.007      &       0.173      &       0.002      &     3.56         & -1.139     \\
$g_{_{F475W}}{-}H$      &       1.042(25.4, 0.144)      &       1.801(6.7, 0.071)      &       1.779(25.9, 0.278)      &       0.005      &       0.009      &       0.001      &     6.69         & -1.266     \\
$g_{_{F475W}}{-}Ks$     &       0.956(33.2, 0.278)      &       2.107(6.6, 0.158)      &       1.705(18.1, 0.116)      &       0.066      &       0.042      &       0.008      &     5.09         & -1.086     \\
$z_{_{F850LP}}{-}J$      &       0.126(32.9, 0.059)      &       0.290(13.7, 0.038)      &       0.479(11.3, 0.054)      &       0.007      &       0.177      &       0.173     &     3.33          & -0.658    \\
$z_{_{F850LP}}{-}H$      &       -0.026(3.7, 0.035)      &       0.148(27.1, 0.084)      &       0.452(27.3, 0.108)      &       0.237      &       0.437      &       0.01      &     2.70          & -1.048    \\
$z_{_{F850LP}}{-}Ks$     &       -0.162(17.2, 0.152)      &       0.058(9.9, 0.044)      &       0.340(30.9, 0.175)      &       0.745      &       0.754      &       0.252     &     1.97          & -0.577    \\
$g{-}J$      &       1.075(27.0, 0.111)      &       1.462(6.8, 0.042)      &       1.807(24.2, 0.167)      &       0.001      &       0.061      &       0.001             &     4.63     &    -1.315      \\
$g{-}H$      &       1.046(26.1, 0.134)      &       1.497(5.4, 0.041)      &       1.884(26.6, 0.162)      &       0.001      &       0.081      &       0.001             &     4.55     &    -1.416      \\
$i{-}J$      &       0.275(27.5, 0.069)      &       0.422(6.3, 0.030)      &       0.666(24.2, 0.119)      &       0.009      &       0.423      &       0.009             &     2.77     &    -1.064      \\
$i{-}H$      &       0.241(25.1, 0.084)      &       0.488(6.0, 0.030)      &       0.745(26.9, 0.119)      &       0.001      &       0.105      &       0.001             &     3.92     &    -1.335      \\
\end{tabular} \\
\tablefoot{Columns list: (1) color; (2-4) first, second and third
  peaks in the three Gaussian model, numbers within parentheses are
  the number of GCs associated with each peak, and the width of the
  distribution; (5-7) GMM $p$-values, as in Table \ref{tab_gmm}; (8)
  separation of the peaks in units of the three Gaussian widths; (9)
  kurtosis of the distribution}
\end{table}
\end{landscape}